\documentclass[aps,12pt,amsmath,amssymb,superscriptaddress,showpacs,showkeys]{revtex4}

\usepackage{dcolumn}
\usepackage{graphicx,color}
\usepackage{textcomp}
\begin {document}
  \newcommand {\nc} {\newcommand}
  \nc {\beq} {\begin{eqnarray}}
  \nc {\eeq} {\nonumber \end{eqnarray}}
  \nc {\eeqn}[1] {\label {#1} \end{eqnarray}}
  \nc {\eol} {\nonumber \\}
  \nc {\eoln}[1] {\label {#1} \\}
  \nc {\ve} [1] {\mbox{\boldmath $#1$}}
  \nc {\ves} [1] {\mbox{\boldmath ${\scriptstyle #1}$}}
  \nc {\mrm} [1] {\mathrm{#1}}
  \nc {\half} {\mbox{$\frac{1}{2}$}}
  \nc {\thal} {\mbox{$\frac{3}{2}$}}
  \nc {\fial} {\mbox{$\frac{5}{2}$}}
  \nc {\la} {\mbox{$\langle$}}
  \nc {\ra} {\mbox{$\rangle$}}
  \nc {\etal} {\emph{et al.\ }}
  \nc {\eq} [1] {(\ref{#1})}
  \nc {\Eq} [1] {Eq.~(\ref{#1})}
  \nc {\Ref} [1] {Ref.~\cite{#1}}
  \nc {\Refc} [2] {Refs.~\cite[#1]{#2}}
  \nc {\Sec} [1] {Sec.~\ref{#1}}
  \nc {\chap} [1] {Chapter~\ref{#1}}
  \nc {\anx} [1] {Appendix~\ref{#1}}
  \nc {\tbl} [1] {Table~\ref{#1}}
  \nc {\fig} [1] {Fig.~\ref{#1}}
  \nc {\ex} [1] {$^{#1}$}
  \nc {\Sch} {Schr\"odinger }
  \nc {\flim} [2] {\mathop{\longrightarrow}\limits_{{#1}\rightarrow{#2}}}
  \nc {\textdegr}{$^{\circ}$}
  \nc {\inred} [1]{\textcolor{red}{#1}}
  \nc {\inblue} [1]{\textcolor{blue}{#1}}
\title{Probing the weakly-bound neutron orbit of $^{31}$Ne with 
total reaction and one-neutron removal cross sections}
\author{W.~Horiuchi}
\email{horiuchi@nt.sc.niigata-u.ac.jp}
\affiliation{Graduate School of Science and Technology, Niigata University, Niigata 950-2181, Japan}
\author{Y.~Suzuki}
\email{suzuki@nt.sc.niigata-u.ac.jp}
\affiliation{Department of Physics, Faculty of Science
and Graduate School of Science and Technology, Niigata University, Niigata 950-2181, Japan}
\author{P.~Capel}
\email{pierre.capel@centraliens.net}
\affiliation{Physique Quantique, C.P. 165/82 and 
Physique Nucl\'eaire Th\'eorique et Physique Math\'ematique, C.P. 229,
Universit\'e Libre de Bruxelles, B 1050 Brussels, Belgium}
\affiliation{National Superconducting Cyclotron Laboratory,
Michigan State University, East Lansing, Michigan 48824, USA}
\author{D.~Baye}
\email{dbaye@ulb.ac.be}
\affiliation{Physique Quantique, C.P. 165/82 and 
Physique Nucl\'eaire Th\'eorique et Physique Math\'ematique, C.P. 229,
Universit\'e Libre de Bruxelles, B 1050 Brussels, Belgium}
\date{\today}
\begin{abstract}
A candidate of a neutron-halo nucleus, $^{31}$Ne, contains a 
single neutron 
in the $pf$ shell. Within the Glauber and eikonal models,
we analyze reactions used to study $^{31}$Ne.
We show in a $^{30}$Ne+n model that 
the magnitudes of the total reaction and above all of the 
one-neutron removal 
cross sections of $^{31}$Ne on $^{12}$C and $^{208}$Pb targets 
strongly depend on the orbital angular momentum of the neutron, 
thereby providing us with efficient ways to 
determine both 
the spin-parity and structure of the ground state of $^{31}$Ne.
Besides these inclusive observables, we also calculate energy
and parallel-momentum distributions for the breakup of $^{31}$Ne,
and show their strong dependence upon the orbital of the valence neutron
in the bound state of $^{31}$Ne.
\end{abstract}
\maketitle
\section{Introduction}
Exploring nuclei near the neutron and proton driplines is making
rapid progresses in and beyond the $p,sd$-shell region.
The Ne isotopes raise interesting structure problems.
The alpha cluster structure 
around $^{20}$Ne is known for many years \cite{HI68,Fuj80}.
Recently $^{17}$Ne, an $^{15}$O+p+p Borromean system, 
has been found to have a large charge 
radius due to a significant amount of $s^2$ 
component \cite{Gei08}.
For the very neutron-rich Ne, Na, and Mg isotopes 
with $N\approx 20$, 
one of the most important issues is 
the vanishing of the shell gap, which causes a mixing of normal 
and intruder configurations, and has significant
influence on the properties of
those nuclei \cite{PR94,CNP98,UOM99,KH01,KH04,Dom06,Doo09}. 
The importance of deformation around $^{30}$Ne is 
stressed in Refs.~\cite{KH01,KH04}, in contrast to the 
result of a mean-field calculation \cite{Sto03}.
The heaviest Ne isotope synthesized so far is $^{34}$Ne.
It may be a dripline nucleus considering  
that $^{33}$Ne is unstable to neutron decay \cite{Not02}.

The nucleus $^{31}$Ne with $N=21$ neutrons
attracts our special attention in view of its 
possible halo structure containing a $1p_{3/2}$ and/or 
$0f_{7/2}$ valence neutron.
Its neutron separation energy $S_{\rm n}$ is 
0.33\,MeV, though it has large uncertainty~\cite{AW03}. 
The ground state spin-parity of $^{31}$Ne is thus 
expected to be either $3/2^-$ or $7/2^-$. 
The former possibility
may happen because the single-particle energy 
of the neutron orbit with low orbital angular 
momentum receives a considerable shift near the neutron
dripline~\cite{HLZ01,Ham07}.
Two calculations, one within a shell model \cite{PR94} and one
using a microscopic cluster model of $^{30}$Ne+n \cite{Des99},
predict that shell inversion.

The rare isotope $^{31}$Ne was first produced 
in a projectile fragmentation reaction~\cite{Sak96}.
Nowadays, an intense beam provided by the Radioactive Ion Beam
Factory (RIBF) at RIKEN 
can produce $^{31}$Ne in sufficiently large amounts
(several particles per second).
Very recently, the total reaction cross sections $\sigma_{\rm
R}$ of heavy Ne isotopes on $^{12}$C target \cite{Oht09pc} and
the one-neutron removal cross sections 
$\sigma_{-\rm n}$ of $^{31}$Ne on $^{12}$C and $^{208}$Pb targets 
have been measured for the first time around 
230~MeV/nucleon \cite{Nak09l}.
The purpose of this article is to analyze the sensitivity
of $\sigma_{\rm R}$, $\sigma_{-\rm n}$ and other 
dissociation cross sections
to the orbit of the \ex{31}Ne valence neutron.
During the completion of this theoretical work,
the one-neutron removal
cross sections of \ex{31}Ne measured at RIKEN became available \cite{Nak09l}.
We seize this opportunity to compare our calculations with the data
to draw conclusions about the structure of the ground state of \ex{31}Ne.

We describe $^{31}$Ne as a system consisting of a $^{30}$Ne 
core ($c$)  and a weakly-bound valence neutron (n).
The core is assumed to be in its $0^+$ ground state though its 
excitation energy is fairly low.
Considering that structure model we evaluate the total reaction 
and one-neutron removal
cross sections within the Glauber formalism \cite{SLY03,AN03,BD04,HSA07}
on both light (\ex{12}C) and heavy (\ex{208}Pb) targets,
and compare the values
obtained for the $1p_{3/2}$ and $0f_{7/2}$ possible configurations
of the \ex{31}Ne ground state.
To predict the sensitivity of more exclusive observables
(e.g. energy and parallel-momentum distributions)
to the ground state configuration, we also perform
calculations within the eikonal model \cite{Glauber,SLY03,AN03,BD04}.
Since both light and heavy targets are considered, we use the
Coulomb correction to the eikonal model (CCE)\cite{MBB03,AS04,CBS08}.

This article is structured as follows: After a summary of
the Glauber and eikonal formalisms (\Sec{th}), we detail the
inputs of our calculations in \Sec{density}.
Our results and analysis are presented in \Sec{res}.
\Sec{conclusion} contains the conclusions and perspectives
of this study.

\section{Theoretical framework}\label{th}
As mentioned in the introduction, we consider in this study
two reaction models.
First, the Glauber model \cite{SLY03,AN03,BD04,HSA07} is used to evaluate
the total reaction and one-neutron removal cross 
sections of \ex{31}Ne.
Second, the eikonal model \cite{Glauber,SLY03,AN03,BD04} is used
to compute the dissociation cross section as a function of the
energy and parallel-momentum between the \ex{30}Ne core and the neutron
after breakup \cite{HBE96,BH04}.
Both models are based on Glauber's idea \cite{Glauber}
to describe the influence of the collision onto
the initial projectile-target wave function
by a multiplying amplitude $e^{i\chi}$,
\beq
\Psi_{\rm f}=e^{i\chi(\ves{b})}\Psi_{\rm i},
\eeqn{eG0}
where the phase $\chi$ is assumed to depend only
on the transverse component $\ve{b}$
of the projectile-target relative coordinate.
In the present work, this phase is obtained
by folding a profile function that describes nucleon-nucleon
effective interactions with the projectile and target densities.
In the eikonal approximation, however, it is more usual to
derive it from optical potentials that simulate the interaction
between the projectile constituents and the target.

In a general interpretation of the eikonal model \cite{SLY03}, the
adiabatic approximation employed in the Glauber model is not assumed,
which invalidates the simple ansatz \eq{eG0} \cite{BCG05}.
The adiabatic approximation ignores the excitation energy compared to
the incident energy, leading to a well-known unphysical result for the
Coulomb dissociation.
In order to solve this problem and still maintain \Eq{eG0},
we only need to correct the Coulomb phase appropriately \cite{MBB03,AS04}.
This approximate version is the CCE and its accuracy has been
tested by comparison to the exact eikonal calculation in \Ref{CBS08}.

In this section, we briefly present both approaches, emphasizing
their common points and differences that make them complementary.

\subsection{Glauber formalism}\label{Glaub}
Provided that $^{31}$Ne can be seen as a neutron loosely bound
to a $^{30}$Ne core whose wave function is 
the same as that of an isolated \ex{30}Ne, $\sigma_{-\rm n}$
can be obtained from the difference between the projectile and
the core interaction cross sections \cite{YOS92,OYS92,SLY03}
\beq
\sigma_{-\rm n}(^{31}{\rm Ne})=\sigma_{\rm I}(^{31}{\rm Ne})
-\sigma_{\rm I}(^{30}{\rm Ne}).
\eeqn{eG1}
Computing the interaction cross sections
is not easy because it excludes inelastic scattering, which
cannot be properly treated if no description of the
internal structure of the projectile is considered.
Fortunately, if the number of bound excited states is small,
$\sigma_{\rm I}$ can be well approximated
by the reaction cross section $\sigma_{\rm R}$, which can be easily
computed within the Glauber formalism \cite{SLY03,AN03,HSA07}.
For \ex{31}Ne, which has only one known bound state, i.e. its 
ground state, this approximation
is legitimate. For \ex{30}Ne, however, $\sigma_{\rm R}(^{30}{\rm Ne})$  will
overestimate $\sigma_{\rm I}(^{30}{\rm Ne})$
by $\sigma_{\rm inel}(^{30}{\rm Ne})$, in which the projectile
is excited towards its $2^+$ and $4^+$ bound excited states.
Nevertheless, the inelastic scattering being a phenomenon occurring 
near the nuclear surface, its contribution
is not expected 
to be significant at incident energies of 200--300~MeV/nucleon where 
the surface transparency becomes large.
The approximation
\beq
\sigma_{-\rm n}(^{31}{\rm Ne})\approx\sigma_{\rm R}(^{31}{\rm Ne})
-\sigma_{\rm R}(^{30}{\rm Ne})
\eeqn{eG2}
seems thus reasonable.

The Glauber model expresses the nuclear part of
the reaction cross section for a nucleus $X$
impinging on a target $T$ as the integral of the
reaction probability with respect to the transverse components $\ve{b}$
of the $X$-$T$ relative coordinate \cite{SLY03,AN03,HSA07}
\beq
\sigma_{\rm R}=\int\left(1-|e^{i\chi(\ves{b})}|^2\right)d\ve{b},
\eeqn{eG3}
where the phase-shift function $\chi$ models the nuclear
interactions between the colliding nuclei.
As mentioned earlier, in the Glauber formalism, this phase
is expressed as a function of the densities of the target
$\rho_T$ and the impinging nucleus $\rho_X$.
It also depends on profile functions $\Gamma_{\rm NN}$
describing effective nuclear interactions between the nucleons.
At the optical limit approximation of the Glauber model (OLA) 
the nuclear phase-shift functions are usually given by \cite{SLY03,AN03,HSA07}
\beq
\chi^N(\ve{b})&=&i \int\!\!\int
\rho_T(\ve{r'})\rho_X(\ve{r''})\Gamma_{\rm NN}(\ve{b}-\ve{s'}+\ve{s''})
d\ve{r''}d\ve{r'},
\eeqn{eG4}
where $\ve{s'}$ and $\ve{s''}$ are the transverse
components of the internal coordinate of the target ($\ve{r'}$)
and the impinging nucleus ($\ve{r''}$), respectively.
The OLA 
is therefore equivalent to the
double-folding of an effective nucleon-nucleon interaction.
Note that the profile functions $\Gamma_{\rm NN}$ depend
on the nucleons considered:
Their expression for identical nucleons (pp or nn) is not
the same as for the proton-neutron (pn) interaction.
Therefore, in our calculations, expression \eq{eG4}
is actually split into 4 terms.
This is done as follows: Replace $\rho_X\Gamma_{\rm
NN}$ with $\rho_X^{\rm p}\Gamma_{\rm Np}+\rho_X^{\rm n}\Gamma_{\rm Nn}$
using the proton and neutron densities of the projectile $X$ 
and change $\rho_T$ by $\rho_T^{\rm p}+\rho_T^{\rm n}$
followed by renaming N of $\Gamma_{\rm NN}$ in accordance with the
density.

As shown by Abu-Ibrahim and Suzuki, the OLA \eq{eG4}
misses some higher-order terms, which can be included using the symmetrized
expression \cite{AS00R,AS00}
\beq
\chi^N(\ve{b})&=&\frac{i}{2}\left( 
\int d\ve{r'}\rho_T(\ve{r'})
\left\{1-\exp\left[-\int d\ve{r''}\rho_X(\ve{r''})
\Gamma_{\rm NN}(\ve{b}+\ve{s'}-\ve{s''})\right]\right\}\right.\nonumber \\
 &+&\left.\int d\ve{r''}\rho_X(\ve{r''})
\left\{1-\exp\left[-\int d\ve{r'}\rho_T(\ve{r'})
\Gamma_{\rm NN}(\ve{b}-\ve{s'}+\ve{s''})\right]\right\}\right).
\eeqn{eG5}
The Glauber calculations presented in the following
are performed using this expression \eq{eG5} in \Eq{eG3}. 
Again the actual phase-shift function in our
calculations is split into four terms. 
The details about the calculation of the densities and the 
profile functions
are summarized in \Sec{density}.

For the carbon target, the Coulomb contribution to the total
reaction cross section is neglected, the reaction 
being fully nuclear dominated.
However, this may no longer be done for heavy targets. 
For the lead target, we add incoherently
to the nuclear reaction cross section \eq{eG3}
the Coulomb contribution at first-order (see \Sec{-n}).

\subsection{Coulomb-corrected eikonal description of reactions}\label{cce}
Since we are also interested in the influence of the \ex{31}Ne
structure on other observables, like energy and parallel-momentum
distributions, we perform calculations
within the eikonal model \cite{Glauber,SLY03,AN03}.
Indeed this model enables us to compute differential cross
sections considering both Coulomb and nuclear interactions,
as their interferences \cite{HBE96,BH04}.
The eikonal model assumes a cluster
structure of the projectile and usually describes the interaction
between the clusters and the target by optical potentials.

\begin{figure}
\includegraphics[width=7cm]{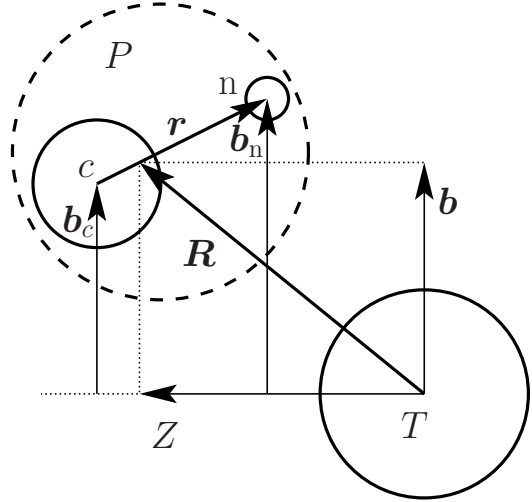}
\caption{Jacobi set of coordinates used for the eikonal calculations:
$\ve{r}$ is the projectile internal coordinate, and
$\ve{R}=\ve{b}+Z\ve{\widehat Z}$ is the target-projectile coordinate.
The transverse components of the core-target and neutron-target coordinates
are denoted by $\ve{b}_c$ and $\ve{b}_{\rm n}$, respectively.}
\label{f0}
\end{figure}

In this work, the projectile $P$ (\ex{31}Ne) is assumed to be made up of a
neutron n of mass $m_{\rm n}$
initially bound to a \ex{30}Ne core $c$ of mass $m_c$ and charge $Z_ce$.
This two-body projectile is impinging on a target $T$ of mass $m_T$
and charge $Z_Te$.
The neutron has spin $I=1/2$, while both core and target
are assumed to be of spin zero.
These three bodies are seen as structureless particles.
\fig{f0} schematizes the set of coordinates we use in the following.
The $c$-n relative coordinate is denoted by $\ve{r}$
and $P$-$T$ relative coordinate by $\ve{R}$,
with $Z$ and $\ve{b}$ its longitudinal and transverse components, respectively.
In \fig{f0}, the transverse parts of the $c$-$T$ ($\ve{b}_c$)
and n-$T$ ($\ve{b}_{\rm n}$) coordinates are shown as well.

The structure of the projectile is
described by the internal Hamiltonian
\beq
H_0=\frac{p^2}{2\mu_{c{\rm n}}}+V_{c{\rm n}}(\ve{r}),
\eeqn{e1}
where
$\ve{p}$ is the relative momentum of the neutron to the core,
$\mu_{c{\rm n}}=m_cm_{\rm n}/m_P$ is the reduced mass of the core-neutron pair
(with $m_P=m_c+m_{\rm n}$),
and $V_{c{\rm n}}$ is the potential describing the core-neutron interaction.
This potential includes a central part,
and a spin-orbit coupling term (see \Sec{density}).

In partial wave $lj$, the eigenstates of $H_0$ are defined by
\beq
H_0 \phi_{ljm}(E,\ve{r})=E \phi_{ljm}(E,\ve{r}),
\eeqn{e2}
where $E$ is the energy of the $c$-n relative motion,
and $j$ is the total angular momentum resulting from the
coupling of the orbital momentum $l$ with the neutron spin $I$.
The negative-energy solutions of \Eq{e2} correspond either to the
physical bound state of the projectile, or to orbitals occupied by the
neutrons of the core, which are forbidden to the valence neutron by the
Pauli principle. 
The former is denoted by $\phi_{l_0j_0m_0}(E_0)$ in the following.
These wave functions are normed to unity.
The positive-energy states describe the broken-up projectile.
Their radial part $r^{-1}u_{klj}$ are normalized according to
\beq
u_{klj}(r)\flim{r}{\infty}\cos\delta_{lj}F_l(kr)+\sin\delta_{lj}G_l(kr),
\eeqn{e3}
where $k=\sqrt{2\mu_{c{\rm n}} E/\hbar^2}$ is the wave number,
$\delta_{lj}$ is the phase shift at energy $E$,
and $F_l$ and $G_l$ are respectively the regular and irregular
Coulomb functions \cite{AS70}.

At the eikonal approximation,
the amplitude appearing in \Eq{eG0}
can be divided into three factors \cite{CBS08}
\beq
e^{i\chi}&=&e^{i\chi_{PT}^C} e^{i\chi^C} e^{i\chi^N},
\eeqn{e7}
where the dependence on the transverse coordinate $\ve{b}$
has been omitted for clarity.
The elastic Coulomb phase $\chi_{PT}^C$
describes the projectile-target Rutherford scattering.
It reads \cite{Glauber}
\beq
\chi_{PT}^C(b)&=& 2\eta\ln(Kb),
\eeqn{e8}
where $K$ is the wave number of the projectile-target relative motion
and $\eta=Z_TZ_ce^2/(4\pi\epsilon_0\hbar v)$ is the $P$-$T$ Sommerfeld
parameter, with $v$ the initial $P$-$T$ relative velocity.

Besides the deflection of the projectile trajectory,
the Coulomb interaction also contributes to the breakup of the projectile.
Acting only on the core, it indeed induces a tidal force
between both components of the projectile.
The Coulomb phase $\chi^C$ in \Eq{e7} simulates that tidal force
(see e.g. Eqs.~(16) and (17) of \Ref{CBS08}).
The slow decrease of this phase at large $b$
leads to divergence in the calculation
of the breakup cross sections \cite{CBS08}.
To overcome this problem, Margueron, Bonaccorso, and Brink have proposed
a correction to this Coulomb term \cite{MBB03}.
It consists in replacing at first order
the Coulomb phase $\chi^C$ by the first order of the perturbation
theory $\chi^{FO}$ (see Eq.~(22) of \Ref{CBS08}) following
\beq
e^{i\chi^C}\rightarrow e^{i\chi^C}-i\chi^C+i\chi^{FO}.
\eeqn{e10}
Because at large $b$ the first-order phase $\chi^{FO}$
decays exponentially, correction \eq{e10} solves the
aforementioned divergence problem.
In addition, it restores most of the missing dynamical effects
in the eikonal model, which enables us to describe reactions
taking on (nearly) the same footing both Coulomb and nuclear
interactions at all orders \cite{CBS08}.

In the eikonal model, the nuclear interactions between the projectile
constituents and the target are usually described by optical potentials
chosen in the literature. In that case, the nuclear phase $\chi^N$
is expressed as integrals over $Z$ of these potentials
\cite{Glauber,SLY03,AN03}.
In the present case, no experimental data exist to constrain such a potential
for the interaction between the \ex{30}Ne core and the target.
Following \Ref{HSA07}, we approximate
the nuclear phase for each projectile constituent
by the OLA \eq{eG4}.
Therefore
\beq
\chi^N(\ve{b},\ve{s})=\chi^N_{cT}(\ve{b}_c)+\chi^N_{{\rm n}T}(\ve{b}_{\rm n}),
\eeqn{e13}
where $\chi^N_{cT}$ and $\chi^N_{{\rm n}T}$ are respectively
the $c$-$T$ and n-$T$ nuclear phases.
They are computed  using \Eq{eG4}, in which the density $\rho_X$
is replaced by the \ex{30}Ne density or a Dirac delta function, respectively.

To evaluate elastic-breakup cross sections within the CCE
we proceed as explained in \Ref{CBS08}.
The elastic-breakup amplitude reads
\beq
S_{kljm}^{(m_0)}(b)=e^{i\left(\sigma_l+\delta_{lj}-l\pi/2+\chi_{PT}^C\right)}
\left\langle\phi_{ljm}(E)\left|
\left(e^{i\chi^C}-i\chi^C+i\chi^{FO}\right)e^{i\chi^N}
\right|\phi_{l_0j_0m_0}(E_0)\right\rangle,
\eeqn{e20}
where $\sigma_l$ is the Coulomb phase shift \cite{AS70}. 

In the following, we consider two breakup observables.
The first is the breakup cross section as a function of the
$c$-n relative energy $E$ after dissociation
\beq
\frac{d\sigma_{\rm bu}}{dE}=\frac{4\mu_{c{\rm n}}}{\hbar^2k}\frac{1}{2j_0+1}
\sum_{m_0}\sum_{ljm}\int_0^\infty b db |S_{kljm}^{(m_0)}(b)|^2.
\eeqn{e21}
The second breakup observable is the parallel-momentum distribution
\beq
\frac{d\sigma_{\rm bu}}{dk_\parallel}=\frac{8\pi}{2j_0+1}
\sum_{m_0}\int_0^\infty b db \int_{|k_\parallel|}^\infty\frac{dk}{k}
\sum_{\nu m}\left|\sum_{lj}(lI m-\nu \nu|jm)
Y_l^{m-\nu}(\theta_k,0)S_{kljm}^{(m_0)}(b)\right|^2,
\eeqn{e23}
where $\theta_k=\arccos (k/k_\parallel)$ is the colatitude of the
$c$-n relative wavevector $\ve{k}$ after breakup.

\section{Densities and potentials}\label{density}
The calculation of the cross sections described in the previous section
requires projectile and target densities and profile functions.
In our study, we follow \Ref{HSA07}.

We first construct \ex{30}Ne densities.
We assume the internal wave function of this nucleus
to be a Slater determinant of single-particle orbitals generated
from the following potential
\begin{equation}
   U(r)=-V_0f(r)+V_1r_0^2\,\ve{l}\cdot\ve{s}
   {\frac{1}{r}}\frac{d}{dr}f(r)+V_C(r)\frac{1-\tau_3}{2},
\label{e30}
\end{equation}
where $\tau_3$ has eigenvalue 1 for neutrons and $-1$ for protons,
and $f$ is the Woods-Saxon form factor
\beq
f(r)=\{1+{\rm exp}[(r-R) /{a}]\}^{-1},
\eeqn{e31}
where radius  $R=r_0A^{1/3}$, with $A=30$.
The spin-orbit strength is set to follow the  systematics \cite{BM1},
\beq
V_1=22-14[(N-Z)/A]\tau_3 
\eeqn{e32}
in MeV. The Coulomb potential $V_C$ is taken from a uniform charge 
distribution. The values of $r_0$ and $a$ are 
varied around 
standard values, and $V_0$ is 
determined separately for neutrons and protons to fit 
$S_{\rm n}$ and $S_{\rm p}$. The resulting values are 
denoted $V_0^{\rm n}$ and $V_0^{\rm p}$, respectively.

The neutron and proton densities of $^{30}$Ne,
$\rho^{\rm n}_c$ and $\rho^{\rm p}_c$, 
are calculated from the occupied orbits by removing 
approximately the effect of the center of mass motion \cite{HSA07}. 
The root mean square (rms) radii for neutron, proton, and matter 
distributions ($r^{\rm n}_c$, $r^{\rm p}_c$, $r^{\rm m}_c$)
are listed in \tbl{t1}. The table also contains 
$\sigma_{{\rm R}}(^{30}{\rm Ne})$ for a 
$^{12}$C target at 100, 240 and 1000\,MeV/nucleon.
The second value of the incident energy is chosen because it is
close to that of the RIKEN experiment \cite{Oht09pc,Nak09l}, and
that profile functions are available at that energy \cite{AHK08}.
The choice of $\Gamma_{\rm NN}$ is explained later in this 
section.

\begin{table}
\begin{center}
\begin{tabular}{ccccccccccccccccccc}
\hline\hline
$r_0$&&$a$&&$V^{\rm n}_0$&&$V^{\rm p}_0$&&
$r^{\rm n}_c$&&
$r^{\rm p}_c$&&
$r^{\rm m}_c$&&&\multicolumn{3}{c}{$\sigma_{\rm R}$}&\\
\cline{15-19}
&&&&&&&&&&&&&&100&&240&&1000\\
\hline	     		     
     &&0.65&&43.71&&72.53&&3.36&&2.58&&3.12&&1.54&&1.29&&1.38\\
1.20 &&0.70&&43.81&&73.55&&3.40&&2.59&&3.16&&1.56&&1.31&&1.39\\
     &&0.75&&43.85&&74.52&&3.46&&2.59&&3.20&&1.59&&1.33&&1.41\\
\hline	     		     
     &&0.65&&41.01&&68.83&&3.44&&2.74&&3.22&&1.58&&1.33&&1.41\\
1.25 &&0.70&&41.15&&69.79&&3.48&&2.73&&3.25&&1.60&&1.34&&1.43\\
     &&0.75&&41.22&&70.72&&3.52&&2.72&&3.28&&1.62&&1.36&&1.44\\
\hline\hline
\end{tabular}
\end{center}
\caption{Single-particle potentials for $^{30}$Ne, rms 
radii of $^{30}$Ne and total 
reaction cross sections of $^{30}$Ne+$^{12}$C collision at the 
incident energy of 100, 240 and 1000\,MeV/nucleon. 
Lengths, energies, and cross sections are given in units of fm, MeV, 
and b.}
\label{t1}
\end{table}

Since \ex{31}Ne is assumed to exhibit a
\ex{30}Ne-n cluster structure, its densities are obtained from the
\ex{30}Ne densities computed above, and the wave function $\phi_{ljm}$
for the \ex{30}Ne-n relative motion.
The latter is determined by solving the
\Sch equation \eq{e2} in either the $1p_{3/2}$ or $0f_{7/2}$ orbital.
The \ex{30}Ne-n interaction is simulated by the same
mean-field potential as for \ex{30}Ne \eq{e30},
but with a different central depth $V_0$.

\begin{figure}
\includegraphics[width=14cm]{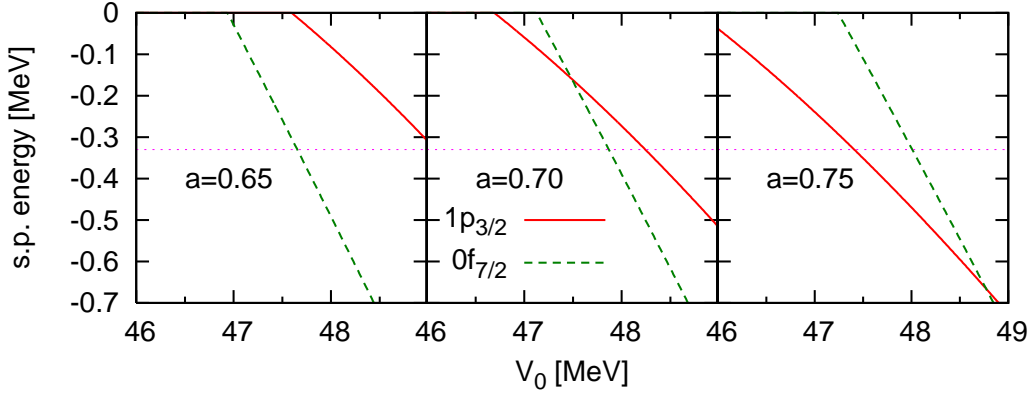}
\caption{(Color online) Single-particle energies of the $1p_{3/2}$ (full lines) and 
$0f_{7/2}$ (dashed lines) valence neutrons of $^{31}$Ne as a 
function of $V_0$ for three different diffuseness parameters $a$ (fm).
The $r_0$ value is 1.25\,fm. Horizontal dotted-line at $-0.33$\,MeV 
denotes the experimental energy. }\label{f1}
\end{figure}

\fig{f1} displays the
single-particle energies of $1p_{3/2}$ and $0f_{7/2}$, 
$\varepsilon(p)$ and $\varepsilon(f)$, 
as a function of $V_0$ for three choices of diffuseness parameter $a$,
the radius parameter being fixed to $r_0=1.25$~fm.
With increasing $a$, $\varepsilon(p)$ 
decreases very rapidly, whereas $\varepsilon(f)$  shows a mild change.
It is therefore possible to obtain the expected shell inversion
by considering a sufficiently large diffuseness (e.g. $a=0.75$~fm).
For actual calculations,
the strength $V_0$ is set to reproduce the $S_{\rm n}$ value of 
0.33 MeV (see \tbl{t2}).
Note that these potentials are also used as $V_{c\rm n}$ in the calculations
of the wave functions $\phi_{ljm}$ that appear
in the eikonal model (see \Sec{cce}).

\begin{figure}
\includegraphics[width=12cm]{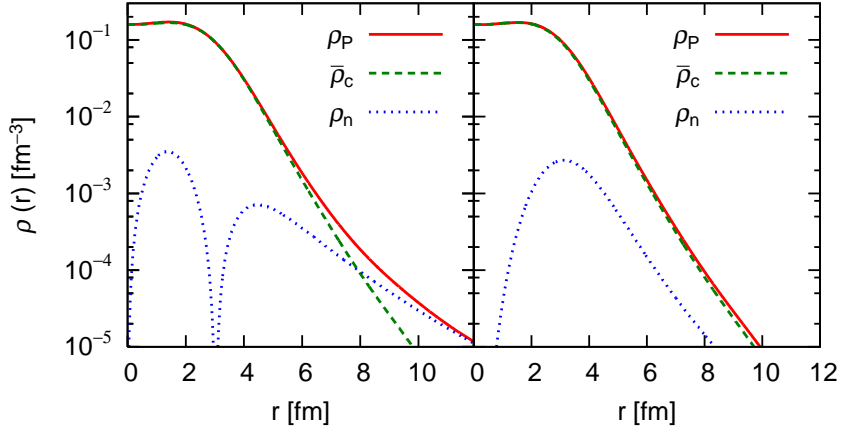}
\caption{(Color online) Matter density of \ex{31}Ne (full lines) and 
contributions from the $^{30}$Ne core (dashed lines) 
and the valence neutron (dotted lines). 
The $r_0$ value is 1.25\,fm, and $a$ is 0.75\,fm for the $1p_{3/2}$ 
orbit (left panel) and 0.70\,fm for the $0f_{7/2}$ orbit (right panel).}
\label{f2}
\end{figure}

The proton and neutron densities 
of $^{31}$Ne, $\rho^{\rm p}_P$ and $\rho^{\rm n}_P$, are
calculated including the recoil effect, which means that
the difference between the centers
of mass of $^{31}$Ne and $^{30}$Ne is treated properly
\beq
\rho^{\rm p}_P(\ve{r'})&=&\overline\rho^{\rm p}_c(\ve{r'})\label{e33a}\\
\rho^{\rm n}_P(\ve{r'})&=&\overline\rho^{\rm n}_c(\ve{r'})+ \rho_{\rm n}(\ve{r'}),
\eeqn{e34a}
where $\ve{r'}$ is the internal coordinate of \ex{31}Ne.
In these expressions, $\overline\rho^{\rm p}_c$ and $\overline\rho^{\rm n}_c$ are
the contributions of the \ex{30}Ne core to the \ex{31}Ne densities.
They slightly differ from the densities of \ex{30}Ne,
because of the recoil effect
\beq
\overline\rho^{\rm p/n}_c(\ve{r'})&=&\frac{1}{2j+1}\sum_m
\int \rho^{\rm p/n}_c(\textstyle{\frac{1}{A+1}}\ve{r}+\ve{r'})
 |\phi_{ljm}(\ve{r})|^2 d\ve{r},
\eeqn{e34c}
where $\ve{r}$ is the \ex{30}Ne-n relative coordinate.
In \Eq{e34a} $\rho_{\rm n}$ denotes the contribution of the valence neutron
to the \ex{31}Ne density
\beq
\rho_{\rm n}(\ve{r'})&=&\frac{1}{2j+1}\sum_m
\int    \delta(\textstyle{\frac{A}{A+1}}\ve{r}-\ve{r'}) 
 |\phi_{ljm}(\ve{r})|^2 d\ve{r}.
\eeqn{e34b}
\fig{f2} displays the matter density of $^{31}$Ne
($\rho_P=\rho^{\rm p}_P+\rho^{\rm n}_P$) as well as its 
contributions from the $^{30}$Ne core
($\overline\rho_c=\overline\rho^{\rm p}_c+\overline\rho^{\rm n}_c$) and 
the valence neutron ($\rho_{\rm n}$). 
The $1p_{3/2}$ orbit (left panel) reaches far in distances and
extends the tail of the $^{30}$Ne density significantly beyond 6 fm.
On the contrary, the $0f_{7/2}$ orbit (right panel)
changes the density only slightly even near the surface.

\tbl{t2} lists the valence-neutron single-particle energies
($\varepsilon$), the rms radii of $^{31}$Ne for the neutron,
proton and matter distributions ($r^{\rm n}_P$, $r^{\rm p}_P$, $r^{\rm m}_P$), and 
$\sigma_{\rm R}(^{31}$Ne) for a $^{12}$C target at 100, 240 and 
1000\,MeV/nucleon.
We also give the rms radius of the valence-neutron orbit
$r_{\rm n}=\sqrt{\langle \ve{r}^2\rangle}$.
This $r_{\rm n}$ value turns out to be around 7\,fm for 
the $1p_{3/2}$ orbit but, due to the larger centrifugal barrier,
is much smaller for the $0f_{7/2}$ orbit: about only 4~fm.
Interestingly, although the matter radii of $^{31}$Ne and $^{30}$Ne depend on 
the potential sets (see Tables~\ref{t1} and \ref{t2}),
their difference remains unchanged:
$\Delta r=r^{\rm m}_P-r^{\rm m}_c$ is 0.19--0.20 for 
$1p_{3/2}$ and 0.04\,fm for $0f_{7/2}$.
The constancy of $\Delta r$ within the set of the 
same $l$ suggests that $\Delta r$ is insensitive 
to the shape of the potential but determined by $S_{\rm n}$ and $l$. 
Despite the fact that the single-particle energy is only $-0.33$\,MeV, 
$\Delta r$ is not very large even for $1p_{3/2}$ because the mass 
number of the core nucleus is fairly large.

Since the neutron separation energy $S_{\rm n}$ of $^{31}$Ne is not
accurately known, we also perform calculations with  
a slightly deeper potential
(see last line of \tbl{t2}) in order to examine the $S_{\rm n}$
dependence of $\sigma_{\rm R}(^{31}{\rm Ne})$ and $\sigma_{-{\rm
n}}(^{31}{\rm Ne})$ values. This potential gives
$\varepsilon(p)=-0.6$~MeV instead of $-0.33$~MeV.  
The matter radius is reduced by only 0.05~fm, but 
the $r_{\rm n}$ value changes by about 0.8~fm. The decrease of  
$\sigma_{\rm R}(^{31}{\rm Ne})$,
and thus of $\sigma_{-{\rm n}}(^{31}{\rm Ne})$, 
on carbon is only about 10~mb. However,
the $\sigma_{-{\rm n}}(^{31}{\rm Ne})$ value on lead is expected to be 
considerably reduced. We will discuss this in \Sec{-n}.

\begin{table}
\begin{center}
\begin{tabular}{ccccccccccccccccccccccc}
\hline\hline
$r_0$&&$a$&&$V_0$&&
$\varepsilon(p)$&&$\varepsilon(f)$&&
$r_{\rm n}$&&
$r^{\rm n}_P$ &&$r^{\rm p}_P$ &&
$r^{\rm m}_P$&&&\multicolumn{3}{c}{$\sigma_{\rm R}$}\\
\cline{19-23}
&&&&&&&&&&&&&&&&&&100&&240&&1000\\
\hline	    	    					 
     &&0.65&&52.28&&$-0.22$&&$-0.33$&&4.25&&3.40&&2.59&&3.16&&1.58&&1.32&&1.41\\
1.20 &&0.70&&51.82&&$-0.33$&&$-0.05$&&7.20&&3.66&&2.60&&3.35&&1.70&&1.41&&1.50\\
     &&0.75&&50.87&&$-0.33$&&--&&7.35&&3.72&&2.60&&3.40&&1.72&&1.43&&1.52\\
\hline	    	    					 
     &&0.65&&47.65&&$-0.01$&&$-0.33$&&4.39&&3.48&&2.74&&3.26&&1.61&&1.35&&1.43\\
1.25 &&0.70&&47.87&&$-0.24$&&$-0.33$&&4.47&&3.52&&2.73&&3.29&&1.63&&1.37&&1.45\\
     &&0.75&&47.41&&$-0.33$&&$-0.07$&&7.44&&3.79&&2.73&&3.48&&1.75&&1.45&&1.55\\
\hline
1.25 &&0.75&&48.52&&$-0.60$&&$-0.55$&&6.62&&3.71&&2.73&&3.43&&1.74&&1.44&&1.54\\
\hline\hline
\end{tabular}
\end{center}
\caption{Properties of the potentials describing $^{31}$Ne.
Using various potential geometries, we adjust
$S_{\rm n}=0.33$~MeV in either
the $1p_{3/2}$ orbit or the $0f_{7/2}$ one.
Last-line potential reproduces $S_{\rm n}=0.60$~MeV
in the $1p_{3/2}$ orbital.
Rms radii of the corresponding densities are listed as well as the
total reaction cross sections of $^{31}$Ne+$^{12}$C at incident 
energies of 100, 240, and 1000 MeV/nucleon.
Lengths, energies, and cross sections are given in units of fm, MeV, and b.}
\label{t2}
\end{table}

The target densities used in our calculations are
obtained from experimental data.
For both \ex{12}C and \ex{208}Pb, the proton densities are derived
from empirical charge densities by removing the finite size 
effect of protons.
The neutron density of \ex{12}C is obtained 
as explained in \Ref{HSA07}.
For \ex{208}Pb, the neutron density is obtained by subtracting 
the proton density from the matter density~\cite{OYS92} taken from
a Hartree-Fock calculation.

Other key inputs to compute the cross sections of \Sec{th} are
the profile functions $\Gamma_{\rm NN}$ that correspond to
effective nucleon-nucleon interactions.
These functions are parametrized in the usual way \cite{HSA07,AHK08}
\beq
\Gamma_{\rm NN}(\ve{b})=\frac{1-i\alpha_{\rm NN}}{4\pi\beta_{\rm NN}}
\sigma^{\rm tot}_{\rm NN} \exp\left(-\frac{b^2}{2\beta_{\rm NN}}\right),
\eeqn{e35}
where $\sigma^{\rm tot}_{\rm NN}$ is the total cross section
for the N-N collision, $\alpha_{\rm NN}$ is the ratio of the real
to the imaginary part of the N-N scattering amplitude,
and $\beta_{\rm NN}$ is the slope parameter of the N-N elastic
differential cross section.
The values of these parameters are taken from 
\Ref{AHK08}. Note that they differ for the 
interaction between identical nucleons (pp or nn)
and for the interaction between a proton and a neutron (pn).
To analyze the sensitivity of our calculations to this choice
of profile functions, we also perform calculations 
that ignore the difference between pp (or nn) and pn interactions.
In those tests, we use the parameters of $\Gamma_{\rm NN}$ 
given in \Ref{HSA07}. 

The profile functions \eq{e35} combined to the densities
of \ex{30,31}Ne and of the target enable us to compute
the phase-shifts \eq{eG5} for the Glauber calculation.
The same parameters are used to derive the OLA 
\eq{eG4} used in the Coulomb-corrected eikonal calculation.
To this end, the densities of the projectile and the target
are expanded on a Gaussian basis
\beq
\rho(r)\approx\sum_i c_i
\exp\Big(-\frac{1}{2}a_ir^2\Big).
\eeqn{e36}
This  enables us to solve analytically
the integrals appearing in Eq.~\eq{eG4} and partly 
in Eq.~\eq{eG5}.
The values $c_i$ and $a_i$ are available from the authors.

In the eikonal model, the nuclear phase \eq{eG4} is added 
to the elastic Coulomb phase \eq{e8} and
the corrected Coulomb phase \eq{e10} to obtain the eikonal phase \eq{e7}.
That phase is then numerically expanded into multipoles
of rank $\lambda$.
To this end, we use a Gauss quadrature on the unit sphere similar to the one
considered to solve the time-dependent \Sch equation in \Ref{CBM03c}.
The number of points along the colatitude is set to
$N_\theta=12$, and the number of points along the azimuthal angle
is $N_\varphi=30$ in most cases but goes up to 40 when large $\lambda$s
are considered.
For the carbon target, the calculation requires a rather large
number of multipoles: $\lambda_{\rm max}=16$
in the $1p_{3/2}$ case, and $\lambda_{\rm max}=12$
in the $0f_{7/2}$ one.
For the lead target, a smaller number of multipoles is needed:
$\lambda_{\rm max}=8$
for the $1p_{3/2}$ state, and $\lambda_{\rm max}=6$
for the $0f_{7/2}$ one.

The eigenfunctions of the projectile Hamiltonian $H_0$ \eq{e2}
are computed numerically with the Numerov method using
1000 radial points equally spaced from $r=0$ up to $r=100$~fm. 
This rather large value is required in order to reach convergence
in the radial integrals appearing in \Eq{e20} and in the calculation
of $r_{\rm n}$, the rms radius of the valence neutron (see \tbl{t2}).
The integrals over $b$ appearing in Eqs.~\eq{e21} and \eq{e23}
are performed numerically from $b=0$ up to $b=400$~fm
with a step $\Delta b=1$~fm. In the $1p_{3/2}$ case this integral
had to be done up to 600~fm to reach convergence when a lead target
was considered.

\section{Discussion of the $1p_{3/2}$ and $0f_{7/2}$ assumptions
within the Glauber model}\label{res}
\subsection{Total reaction cross sections}\label{trcs}

\begin{figure}
\includegraphics[width=10cm]{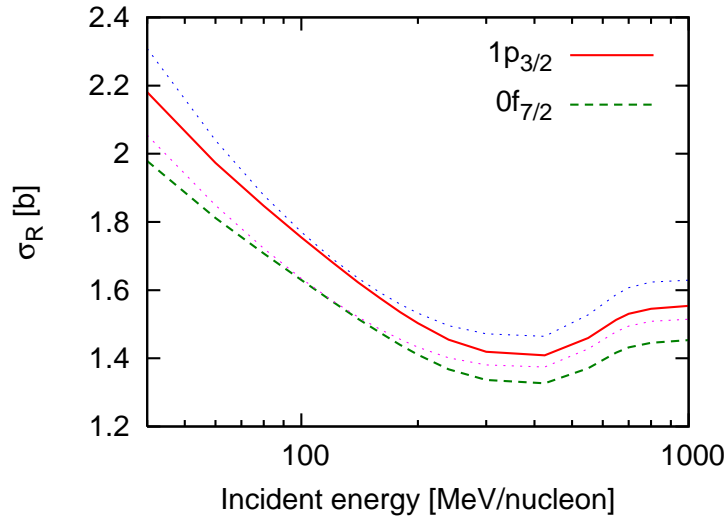}
\caption{(Color online) Total reaction cross section of \ex{31}Ne
on a \ex{12}C target as a function of incident energy. Dotted lines 
are the results with the OLA phase shifts \eq{eG4}.
See caption of \fig{f2} for $r_0$ and $a$.}
\label{fadd1}
\end{figure}

\fig{fadd1} compares $\sigma_{\rm R}(^{31}{\rm Ne})$ 
on a $^{12}$C target calculated
within the Glauber model (see \Sec{Glaub}) 
for the $1p_{3/2}$ (full line) 
and $0f_{7/2}$ (dashed line) orbits as a function
of the \ex{31}Ne incident energy. 
The phase-shift function is calculated using \Eq{eG5}.
The projectile density is obtained using the potential sets
of radius $r_0=1.25$~fm with diffuseness $a=0.75$~fm for the $1p_{3/2}$ orbit
and $a=0.70$~fm for the $0f_{7/2}$ orbit.
At all energies the relative difference in $\sigma_{\rm R}$ between both
configurations is about 5--10\%. 
For example, as listed in Table~\ref{t2}, 
$\sigma_{\rm R}(^{31}{\rm Ne})$ at 240~MeV/nucleon is 
1.45~b for the $p$ orbit and 1.37~b for the $f$ one.
Thus the difference of $\sigma_{\rm R}(^{31}{\rm Ne})$
depending on whether the 
orbital angular momentum of the valence neutron is 
1 or 3 amounts to 87~mb. 
Though not very large, this difference may be sufficient 
to determine which assignment is favorable in comparison 
with experiment~\cite{Oht09pc}.

The reaction cross section
is larger for a $1p_{3/2}$ neutron than for the $0f_{7/2}$ neutron
because the integral appearing in the
phase shifts \eq{eG5} extends on a larger domain
in the former case than in the latter.
This variation in $\sigma_{\rm R}(^{31}{\rm Ne})$
with the projectile configuration,
being mostly due to the change
in the valence-neutron orbital
is therefore rather small. 
Indeed, most of $\sigma_{\rm R}(^{31}{\rm Ne})$ is contributed by
the $^{30}$Ne core, whose reaction cross section
does not vary much with the potential set:
$\sigma_{\rm R}(^{30}{\rm Ne})=1.36$~b for $r_0=1.25$~fm and $a=0.75$~fm,
and 1.34~b for $r_0=1.25$~fm and $a=0.70$~fm.
On the contrary, the increase in 
$\sigma_{\rm R}$ due to the addition of the valence neutron, 
$\sigma_{\rm R}(^{31}{\rm Ne})-\sigma_{\rm R}(^{30}{\rm Ne})$, 
is strongly  dependent on the assumed configuration: The increase
turns out to be 96~mb for the $1p_{3/2}$ orbit 
and 26~mb for the $0f_{7/2}$ orbit at 240~MeV/nucleon.
Following \Eq{eG2}, this result suggests the one-neutron removal
cross section to be an observable more sensitive to the projectile
configuration (see \Sec{-n}).

To investigate the sensitivity of our calculations to the
construction of the phase-shift function, we also
compute $\sigma_{\rm R}(^{31}{\rm Ne})$
using the OLA \eq{eG4} (dotted lines in \fig{fadd1}).
As is usually observed~\cite{HSA07,AHK08}, 
the OLA tends to predict larger cross sections.
However, the difference between the reaction cross sections
obtained with the $1p_{3/2}$ configuration and the $0f_{7/2}$ one
is about the same using OLA \eq{eG4}
as when the phase-shift function~\eq{eG5} is used.

\begin{figure}
\includegraphics[width=10cm]{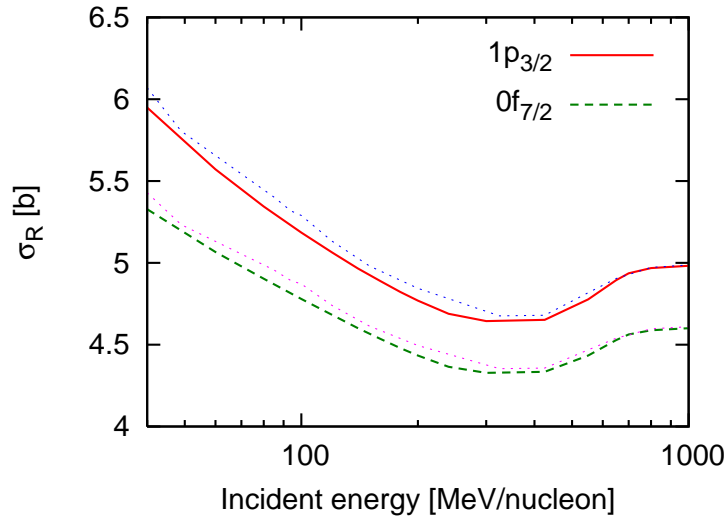}
\caption{(Color online) Nuclear contribution to the total 
reaction cross section of \ex{31}Ne
on a $^{208}$Pb target as a function of incident energy. The dotted
 lines are the results obtained with the average profile functions 
taken from \Ref{HSA07}.
See caption of \fig{f2} for $r_0$ and $a$.}
\label{fadd2}
\end{figure}

\fig{fadd2} displays $\sigma_{\rm R}(^{31}{\rm Ne})$ on 
a $^{208}$Pb 
target calculated with only the nuclear phase shifts.
The effect of Coulomb breakup is discussed in the next subsection. 
As observed for the carbon target, the difference between the
$1p_{3/2}$ (full line) and $0f_{7/2}$ (dashed line)
configurations is small though non-negligible.
As mentioned earlier, this difference comes mainly
from the valence-neutron contribution.
The increase of $\sigma_{\rm R}$ from $^{30}$Ne to $^{31}$Ne 
is even more striking for a $^{208}$Pb target.
It is almost ten times larger considering a $1p_{3/2}$ valence neutron
than a $0f_{7/2}$ one.
At 240~MeV/nucleon,
the reaction cross section increases from 4.36~b to 4.69~b in the former case
while it goes from 4.33~b to only 4.37~b in the latter.

Since the proton and neutron densities of the lead target are 
different, we examine how much the cross sections 
depend on the choice of the 
profile function $\Gamma_{\rm NN}$. \fig{fadd2} compares 
two sets of calculations, one which employs different interactions 
between pp (or nn) and pn (full and dashed lines),
and the other which uses the averaged interaction 
taken from \Ref{HSA07} (dotted lines).
As observed in \fig{fadd2},
the choice of the averaged
interaction tends to slightly overestimate the cross sections below 300
MeV/nucleon.

The enhanced cross section for the $1p_{3/2}$ orbit 
reflects the spatial extension of the neutron orbit. If its 
$S_{\rm n}$ value is increased to, say 0.6~MeV as shown in 
\tbl{t2}, $\sigma_{\rm R}(^{31}{\rm Ne})$ gets smaller compared to 
that with $S_{\rm n}=0.33$~MeV: 
At 240~MeV/nucleon, it is reduced by 12~mb 
for carbon and by 65~mb for lead. These cross sections are still 
significantly larger than those for 
the $0f_{7/2}$ neutron case.

\subsection{One-neutron removal cross sections}\label{-n} 
\begin{figure}
\includegraphics[width=10cm]{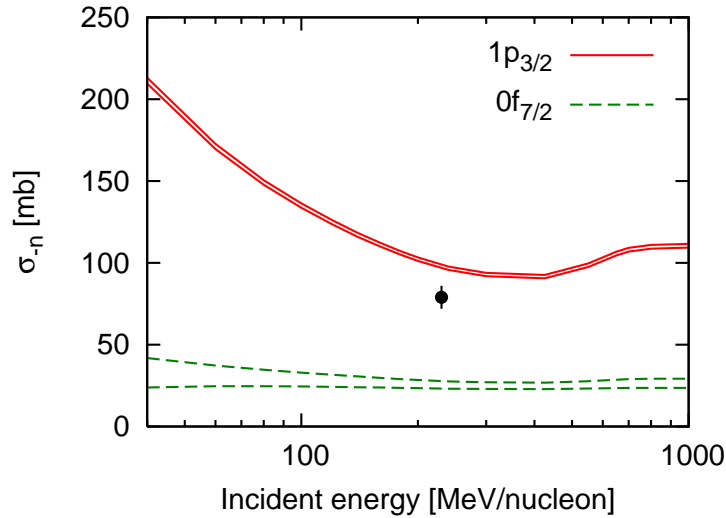}
\caption{(Color online) One-neutron removal cross section of \ex{31}Ne,
approximated by
$\sigma_{\rm R}(^{31}{\rm Ne})-\sigma_{\rm R}(^{30}{\rm Ne})$,
on a \ex{12}C target as a function of incident energy: 
$1p_{3/2}$ (range between full lines) and $0f_{7/2}$ (range
 between dashed lines).
The experimental point is from \Ref{Nak09l}.}
\label{f3}
\end{figure}

As mentioned in \Sec{Glaub}, we
evaluate the one-neutron removal cross section $\sigma_{-\rm n}$
for \ex{31}Ne on carbon and lead targets using approximation \eq{eG2}. 
\fig{f3} shows the results obtained on a \ex{12}C target
as a function of the \ex{31}Ne incident energy
for both $1p_{3/2}$ (full lines) and
$0f_{7/2}$ (dashed lines) configurations.
To evaluate the sensitivity of these results to the
potential set used to generate the projectile densities, we have
performed the calculations with the different potentials given in
Tables~\ref{t1} and \ref{t2}.
Though the $1p_{3/2}$ or $0f_{7/2}$ orbits vary with
the potential set, they predict very similar
$\sigma_{-\rm n}$ values: In both cases 
these values are contained
between the pairs of lines shown in \fig{f3}. Hereafter 
we use the potential set with $r_0=1.25$ fm and $a=0.75$ fm for 
the $1p_{3/2}$ orbit, and the set with $r_0=1.25$ fm and $a=0.70$ fm 
for the $0f_{7/2}$ orbit unless otherwise mentioned.

As discussed in the previous subsection, the
interesting result of this set of calculations is that
$\sigma_{-\rm n}$ is always much larger for a $1p_{3/2}$ valence neutron
than for a $0f_{7/2}$ one.
At 240~MeV/nucleon, close to the energy of the 
RIKEN experiment \cite{Nak09l},
the former configuration leads to a cross section of about 
96~mb, whereas the latter gives only 26~mb.
This difference is basically due to the larger spatial extension of the
$p$ orbit compared to that of the $f$ orbit, which is due to
the change in the centrifugal barrier.
The experimental cross section amounts to 79(7)~mb \cite{Nak09l}.
This value, being both close to our $1p_{3/2}$ calculation
and much higher than our $0f_{7/2}$ one, favors
a ground state wave function for \ex{31}Ne strongly dominated
by a configuration in which the valence neutron is in the
$1p_{3/2}$ orbital coupled to a
\ex{30}Ne core in its $0^+$ ground state.
This comparison therefore suggests a $3/2^-$ spin-parity
for the \ex{31}Ne ground state,
rather than the $7/2^-$ deduced from the naive shell model.

As shown in \fig{f3}, the difference in the magnitude of
$\sigma_{-\rm n}$ increases at lower incident energies.
An experiment performed at such an energy (e.g. a few tens of MeV/nucleon)
would improve the confidence in the identification of
the \ex{31}Ne configuration.

\begin{figure}
\includegraphics[width=10cm]{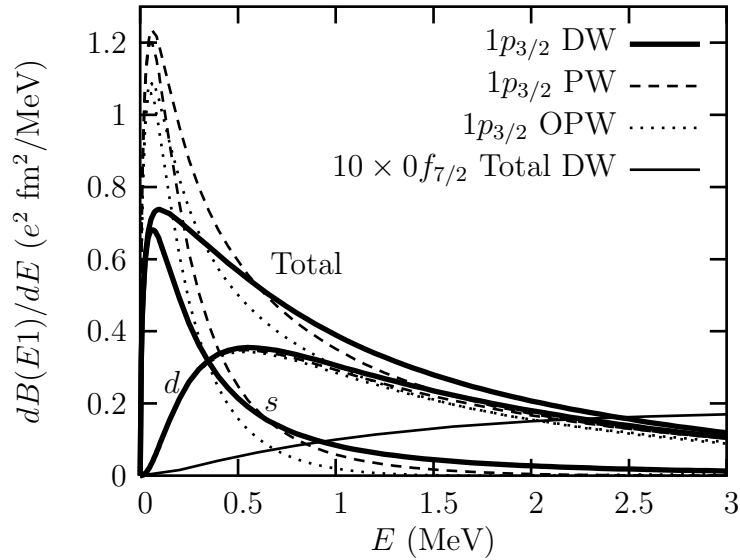}
\caption{Electric dipole transition strength of $^{31}$Ne
as a function of the $c$-n relative energy $E$ after dissociation.
For the $1p_{3/2}$ transition, both contributions from  
the $s$- and $d$-wave continuum states are shown.
Results obtained with distorted waves (DW, full lines),
plane waves (PW, dashed lines),
and orthogonalized plane waves (OPW, dotted lines)
are shown separately.}
\label{f4}
\end{figure}

To evaluate $\sigma_{-\rm n}$ for a $^{208}$Pb target
we may no longer neglect the Coulomb contribution to the
one-neutron removal process. 
Since the Coulomb interaction contributes
mostly to the elastic breakup, we add an estimate
of the Coulomb-breakup cross section 
to the reaction cross section
computed within the Glauber framework.
To this end, we use the first-order
of the perturbation theory, considering only the dominant
dipole transition.
In that approximation, the $1p_{3/2}$ neutron is excited to
continuum states with $l=0$ or 2, whereas the $0f_{7/2}$ neutron 
is moved to $d$ or $g$ positive-energy states.
This Coulomb contribution to $\sigma_{-\rm n}$ can be estimated 
by integrating the electric dipole transition strength $dB(E1)/dE$
multiplied by the photon number spectrum over 
the excitation energy \cite{SLY03}.
\fig{f4} compares the $dB(E1)/dE$ distributions for the initial
$p$ (thick full lines) and $f$ orbits (thin full line).
In the former case, the partial-wave contributions are shown as well.
Note that the result obtained from the initial $0f_{7/2}$ configuration
is multiplied by 10 for readability.
These quantities depend on the 
choice of a minimum impact parameter $b_{\rm {min}}$ from which 
the Coulomb breakup is assumed to contribute.
However, the dependence of $\sigma_{-\rm n}$  
on $b_{\rm {min}}$ is found to be moderate around $b_{\rm min}=12.7$~fm,
which is obtained from $b_{\rm min}=r_{\rm eff}(31^{1/3}+208^{1/3})$,
with $r_{\rm eff}=1.4$~fm.

The dipole strength obtained for the $1p_{3/2}$ configuration 
is concentrated at low excitation energy.
The $s$ wave gives a larger contribution to that distribution
than the $d$ wave at $E< 0.5$\,MeV, but the $d$ wave dominates over the 
$s$ wave with increasing energy.
On the contrary, $dB(E1)/dE$ for the $0f_{7/2}$ initial state,
besides being much smaller than the
$1p_{3/2}$ one, has a completely different energy dependence:
It is flat and extends to high energies.
This suggests that differential observables,
like energy or parallel-momentum distributions,
could be used to discriminate between these two possible configurations
(see \Sec{distr}).

To evaluate the sensitivity of this calculation to the
$c$-n final state interactions, we evaluate the dipole strength for the
initial $1p_{3/2}$ bound state using
distorted waves (DW, i.e., positive-energy eigenstates of the $c$-n
Hamiltonian \eq{e2}; full lines),
plane waves (PW; dashed lines),
or orthogonalized plane waves (OPW, i.e., plane waves
orthogonalized to the
Pauli-forbidden bound states of Hamiltonian \eq{e2} \cite{AS04};
dotted lines).
Interestingly only the $s$ wave contribution is sensitive to the
continuum description: That value is much reduced in the vicinity
of its maximum when DW are considered instead of PW or OPW.
Nevertheless, these changes do not affect the results as
much as to modify our conclusions.

At 240~MeV/nucleon, and using DW, we obtain 0.81~b for 
the Coulomb contribution to $\sigma_{-\rm n}$:
0.32~b from the $s$ wave and 0.49~b from the $d$ waves.
This value is added incoherently to the nuclear
contribution to $\sigma_{-\rm n}$, which is 
estimated to be about 0.33~b in the Glauber model.
The resulting $\sigma_{-\rm n}$ value turns out to be 1.14~b.
As expected, the dipole strength obtained for the $f$ orbit
is much smaller: Its contribution to $\sigma_{-\rm n}$ is a mere
57~mb. The nuclear contribution 
is evaluated in the Glauber model to be about 34~mb,  
leading to a total $\sigma_{-\rm n}=91$~mb. 
This is about one order of magnitude smaller than 
the cross section for the $p$ orbit.
The experiment performed at RIKEN gave $\sigma_{-\rm n}=712(65)$~mb
\cite{Nak09l}.
Thus again slightly below our theoretical prediction for the
$1p_{3/2}$ configuration, and much higher than the cross
section obtained for the $0f_{7/2}$ orbit.
This confirms the shell inversion predicted by former structure calculations
\cite{PR94,Des99}, in agreement with the analysis of
Nakamura \etal \cite{Nak09l}.
Note that evaluations of the Coulomb contribution
using PW or OPW lead to similar results: large $\sigma_{-\rm n}$ for
the $1p_{3/2}$ configuration,
and small $\sigma_{-\rm n}$ for the $0f_{7/2}$ one.

As mentioned in the last paragraph of the previous 
subsection, the Coulomb breakup contribution will be very sensitive 
to $S_{\rm n}$ of the $1p_{3/2}$ orbit. We have repeated the 
calculation assuming $S_{\rm n}=0.6$~MeV.
The $\sigma_{-\rm n}$ value for $S_{\rm n}=0.6$~MeV 
is predicted to be 0.75~b, of which 0.49~b is due to
the Coulomb breakup.
Changing $S_{\rm n}$ from 0.33~MeV to 0.6~MeV 
thus reduces $\sigma_{-\rm n}$ by 0.32~b.
This is much larger than the corresponding 
reduction (65~mb) in the nuclear breakup contribution.
Since $\sigma_{-\rm n}$ changes 
significantly as a function of $S_{\rm n}$ mainly because of 
the Coulomb dissociation, a
close analysis of $\sigma_{-\rm n}$ on a $^{208}$Pb target 
can give some constraint on the $S_{\rm n}$ value of $^{31}$Ne.
The one-neutron removal cross section obtained
with $S_{\rm n}=0.6$~MeV being closer to the experimental value,
suggests that the one-neutron separation energy of \ex{31}Ne might
be higher than 0.33~MeV. However, this reduction from theory
to experiment may also be due to a spectroscopic
factor for the $1p_{3/2}$ configuration smaller than one.
Other observables, like energy or parallel-momentum distributions
for elastic breakup, may provide further valuable information.

\section{Eikonal calculation of energy and parallel-momentum distributions}\label{distr}
Besides the significant difference in magnitude between
the one-neutron removal cross section, the distinction between
the $1p_{3/2}$ and $0f_{7/2}$ configurations for \ex{31}Ne
could be made by looking at differential breakup observables,
like energy or parallel-momentum distributions.
To analyze the influence of the \ex{31}Ne configuration on such
cross sections, we perform elastic-breakup calculations within the
Coulomb-corrected eikonal model (CCE, see \Sec{cce} and \Ref{CBS08}).
Unlike the Glauber model, the CCE solves the divergence
problem posed by the Coulomb interaction between
the projectile and the target.
This enables us to take account of
nuclear and Coulomb interactions on the same footing and to include
their interference in the description of the reaction process.
The following calculations are performed with the inputs
detailed in \Sec{density}.

\begin{figure}
\includegraphics[width=10cm]{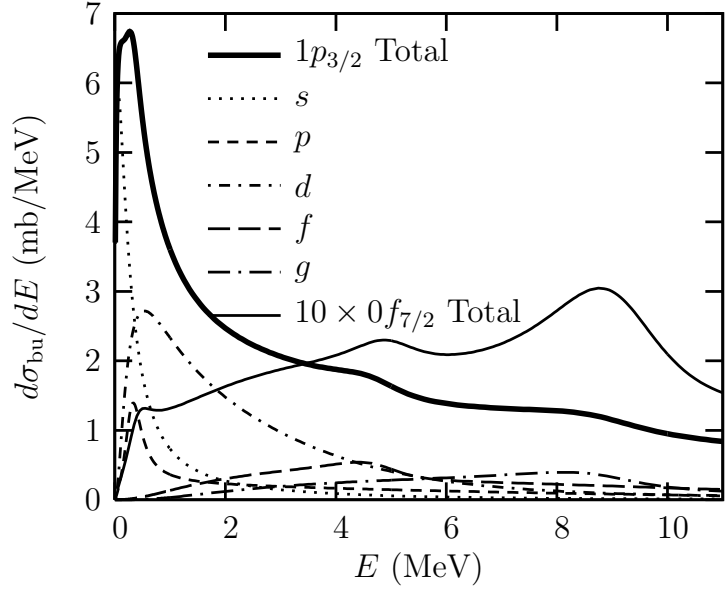}
\caption{Energy distribution for the elastic breakup of \ex{31}Ne on carbon
at 240~MeV/nucleon. The total cross section for the initial
$1p_{3/2}$ bound state is plotted as well as its major
partial-wave contributions.
The result obtained with the initial $0f_{7/2}$ bound state, multiplied by 10,
is shown for comparison.}
\label{f5}
\end{figure}

The elastic-breakup cross sections obtained for \ex{31}Ne impinging on a
carbon target at 240~MeV/nucleon are shown in \fig{f5}
as a function of the energy $E$
between the \ex{30}Ne core and the neutron after dissociation.
The total cross section for the $1p_{3/2}$ configuration 
is displayed with the thick full line, while
its dominant $s$--$g$ contributions are plotted with interrupted lines.
The breakup cross section obtained considering the $0f_{7/2}$ ground state
is depicted with the thin full line.
Note that it is multiplied by 10 for readability.
Both distributions differ significantly.
First, as already mentioned in \Sec{-n}, the magnitude
of the $0f_{7/2}$ cross section is much lower than the $1p_{3/2}$ one.
Second, the $1p_{3/2}$ distribution is strongly peaked at low energy,
whereas the $0f_{7/2}$ distribution extends over a broader energy domain.
This confirms that in addition to one-neutron removal cross sections,
energy distributions could be used to determine the configuration
of \ex{31}Ne ground state.

The two bumps observed in the $0f_{7/2}$ cross section at about 5 and 9~MeV
correspond to $f_{5/2}$ and $g_{9/2}$ resonances
of widths $\Gamma_{0f_{5/2}}\simeq 1.5$~MeV and
$\Gamma_{0g_{9/2}}\simeq 3$~MeV, respectively.
These resonances are produced by the $c$-n potential
used in this calculation (see \tbl{t2}),
but were not fitted to any known state.
In the present work they have thus no physical meaning.
However, this result indicates that if \ex{31}Ne were to exhibit
resonant states with a strong \ex{30}Ne-n cluster structure,
these could be revealed by a measurement of
elastic breakup on a light target \cite{CGB04}.

\begin{figure}
\includegraphics[width=10cm]{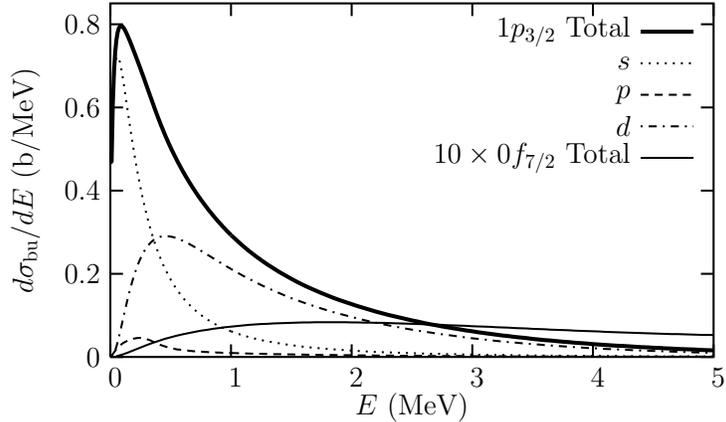}
\caption{Same as \fig{f5} for a lead target.
Note the change in the energy axis.}\label{f6}
\end{figure}

These resonances are also present in the $1p_{3/2}$ calculation,
but the bumps they generate are less marked than in the $0f_{7/2}$ case.
The $1p_{3/2}$ orbit, being two quanta of orbital angular momentum
further away from the resonances than the $0f_{7/2}$ state,
is indeed less prone to be excited towards that part of the continuum.

We also perform a similar calculation for a \ex{208}Pb target.
The corresponding energy distributions are plotted in \fig{f6}.
As in the nuclear breakup case, the two configurations lead
to very different results. Not only is the magnitude of the
distribution strongly dependent on the initial state
(note that the $0f_{7/2}$ cross section is multiplied by 10),
but also its shape clearly reveals the configuration of \ex{31}Ne
ground state.
As in Figs.~\ref{f5} and \ref{f4},
the $1p_{3/2}$ energy distribution is peaked at
low energy and decreases rapidly with $E$.
The $0f_{7/2}$ cross section, on the contrary, is much flatter.

\begin{figure}
\includegraphics[width=10cm]{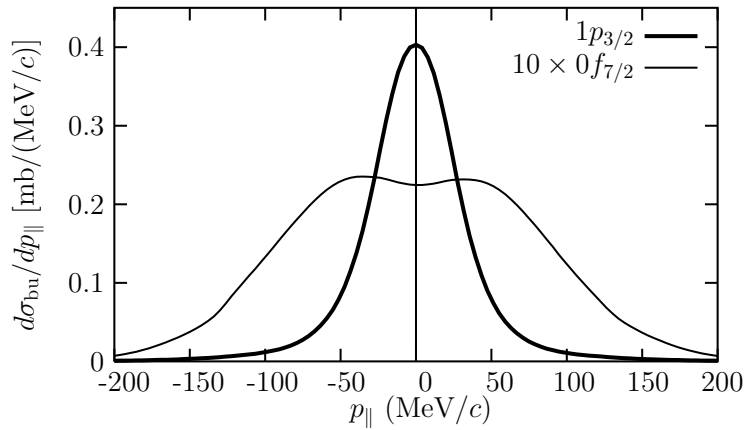}
\caption{Parallel-momentum distribution for the elastic breakup
of \ex{31}Ne on carbon
at 240~MeV/nucleon. The cross section for the initial
$1p_{3/2}$ bound state is compared to that
obtained with the initial $0f_{7/2}$ bound state.
The latter is multiplied by 10.}\label{f7}
\end{figure}

Another observable that is often used to discriminate
the orbital of valence nucleons is the parallel-momentum
distribution \cite{Han96,Sau00,HT03}.
In that case, the breakup cross section is evaluated
as a function of the parallel-momentum between the core and
the neutron after dissociation.
\fig{f7} depicts the parallel-momentum distribution
for the elastic breakup of \ex{31}Ne on a carbon target at 240~MeV/nucleon.
The results obtained with both the $1p_{3/2}$ (thick line) and
$0f_{7/2}$ (thin line) configurations are shown. Note that here also
the latter is multiplied by 10 for clarity.

The signature of the initial configuration is even clearer here
than in the energy distribution. Besides the significant change
in magnitude, we observe that the $0f_{7/2}$ parallel-momentum distribution
is much broader than that of the $1p_{3/2}$ configuration.
This distribution can be understood as a reminiscence of the initial
bound-state wave function expressed in the momentum space
\cite{Han96,HT03}. 
The large spatial expansion of the $1p_{3/2}$ wave function
translates into a narrow momentum distribution, which is
revealed in this breakup cross section.
On the contrary, the narrower spatial distribution of the
$0f_{7/2}$ state leads to the broader parallel-momentum distribution
observed in \fig{f7}.

\section{Conclusion and perspectives}\label{conclusion}
The very neutron-rich isotope \ex{31}Ne ($N=21$) is located in a region
where mixing of normal and intruder shell configurations is expected.
In a naive shell model, the \ex{31}Ne ground state
would thus be seen as a \ex{30}Ne core in its $0^+$ ground state
to which a $0f_{7/2}$ valence neutron is added.
However, some calculations predict this valence neutron to be in a $1p_{3/2}$
intruder orbital instead \cite{PR94,Des99}.
If this were the case, the low angular momentum of the orbital combined to the
low one-neutron separation energy of \ex{31}Ne
($S_{\rm n}\simeq 0.33$~MeV \cite{AW03})
would suggest this nucleus to exhibit a one-neutron halo.

Recently, the new RIBF facility at RIKEN has produced a \ex{31}Ne
beam at about 230~MeV/nucleon.
This beam is sufficiently intense to allow the measurement of
its total reaction and one-neutron removal cross sections
on carbon and lead targets \cite{Oht09pc,Nak09l}.
The present work aims at analyzing the sensitivity of these cross sections
to the structure of the exotic isotope \ex{31}Ne.
To this aim we use the Glauber model detailed
in \Ref{HSA07} to evaluate $\sigma_{\rm R}$ and
$\sigma_{-\rm n}$.
This theoretical work shows that 
both $\sigma_{\rm R}$ and $\sigma_{-\rm n}$
computed considering a $1p_{3/2}$ configuration
for \ex{31}Ne are larger than those obtained with
a $0f_{7/2}$ valence neutron.
Especially, the difference in $\sigma_{-\rm n}$
is significant enough to doubtlessly discriminate between the
two possible configurations.
During the completion of this theoretical work,
the one-neutron removal cross sections of \ex{31}Ne
measured at RIKEN became available \cite{Nak09l}.
The comparison of these data to our calculations
suggests a strong $1p_{3/2}$ configuration in the
wave function of \ex{31}Ne ground state, confirming,
independently from the analysis of Nakamura \etal \cite{Nak09l},
the expected shell inversion in \ex{31}Ne.
We therefore conclude the spin-parity of that ground state
to be $3/2^-$ rather than $7/2^-$ as suggested by the naive shell model.

Since other observables could be used to test this
shell inversion, we have also performed prospective
calculations within the Coulomb-corrected eikonal
approximation \cite{CBS08} for the breakup of \ex{31}Ne
on both carbon and lead targets.
These calculations confirm that a $0f_{7/2}$ configuration
would lead to much smaller breakup cross sections than if the valence neutron
were in the intruder $1p_{3/2}$ orbital.
They also show that the shape of the energy and parallel-momentum
distributions could be used to distinguish between the two
possible configurations. Indeed, whereas assuming a $1p_{3/2}$ valence
neutron gives energy distributions peaked at low energy,
the $0f_{7/2}$ configuration leads to distributions that
reach much higher energies.
We have also observed that the parallel-momentum
distribution is much narrower when the bound state is assumed
in the $p$ partial wave than in the $f$ one.
The measurement of these distributions
would therefore provide a complimentary way to confirm the
structure information obtained from the recent RIKEN measurement
of $\sigma_{-\rm n}$.

At such a distance from the valley of stability 
and near the region of the island of inversion, 
the \ex{31}Ne ground state may not be composed
of a single configuration. An extension of the reaction models used
in this work to a multiple-configuration description
of the projectile structure,
as the one proposed by Summers \etal \cite{SNT06},
would definitely improve the reaction model.
Such a model would indeed help understanding
the influence of a multiple-configuration structure of the projectile
upon reaction observables.

\begin{acknowledgments}
This work has been done in the framework of the agreement between the
Japan Society for the Promotion of Science (JSPS)
and the Fund for Scientific Research of Belgium (F. R. S.-FNRS).
W.~H acknowledges a support by a Grant-in Aid for 
Scientific Research for Young Scientists (No. 19$\cdot$3978). 
Y.~S is supported by a Grant-in Aid for 
Scientific Research (No. 21540261).
D.~B. acknowledges travel support of the
Fonds de la Recherche Scientifique Collective (FRSC).
P.~C. acknowledges the support of the F. R. S.-FNRS
and of the National Science Foundation grant PHY-0800026.
This text presents research results of the Belgian program P6/23 on
interuniversity attraction poles initiated by the Belgian-state
Federal Services for Scientific, Technical and Cultural Affairs (FSTC).
\end{acknowledgments}

\end{document}